\documentstyle[12pt,aaspp4,psfig]{article}
\def\etal{{\it et al.\/}}

\begin{document}

\title{\bf The Cluster of Galaxies Surrounding Cygnus A}

\author{Frazer N. Owen$^1$}
\affil{National Radio Astronomy Observatory$^2$ \\ 
Socorro, New Mexico  87801 \\
fowen@aoc.nrao.edu}
\vskip 0.2in
\author{Michael J. Ledlow$^1$}
\affil{New Mexico State University, Dept. of Astronomy \\ 
Las Cruces, NM 88003 \\
mledlow@nmsu.edu}

\vskip 0.2in
\author{Glenn E. Morrison$^1$,$^3$}
\affil{National Radio Astronomy Observatory$^{2}$ \\
Socorro, New Mexico 87801
\\ gmorriso@aoc.nrao.edu} 
\centerline{and}
\vskip 0.2in
\author{John M. Hill}
\affil{Steward Observatory, University of Arizona 
\\ Tucson, AZ 85721
\\ jhill@as.arizona.edu}
\vskip 0.2in
\baselineskip 10pt 
$^1$Visiting Astronomer, Kitt Peak National Observatory, National Optical 
Astronomy Observatories, operated by the Association of Universities for 
Research in Astronomy, Inc., under contract with the National Science 
Foundation. 

$^2$The National Radio
Astronomy Observatory is operated by  Associated Universities, Inc., under
a cooperative agreement with the National Science Foundation.

$^3$ Also University of New Mexico, Dept. of Physics \& Astronomy, Albuquerque,
NM 87131

\clearpage

\begin{abstract} 
\baselineskip 24pt
	We report optical imaging and spectroscopy of 41 galaxies in
a 22 arcmin square region surrounding Cygnus A. The results show that
there is an extensive rich cluster associated with Cygnus A of Abell richness
at least 1 and possibly as high as 4. The velocity
histogram has two peaks, one centered on Cygnus A, and a more significant peak
redshifted by about
2060 km s$^{-1}$ from the velocity of Cygnus A. The dynamical centroid of the 
spatial
distribution is also shifted somewhat to the NW. However, statistical
tests show only weak evidence that there are two distinct clusters. The
entire system has a velocity dispersion of 1581 km s$^{-1}$
which is slightly larger than other, well studied, examples of
rich clusters. 


\end{abstract}

\clearpage 
\section{Introduction}

	The well-known, radio luminous, prototypical FR II radio
galaxy, Cygnus A has been known for some time to be embedded in
an extensive, cluster-like cloud of hot gas. The gas distribution was first imaged
well by the {\it Einstein} x-ray satellite (Arnaud \etal \markcite{arnaud84}
1984). The best high resolution (ROSAT HRI) 
image was made by Carilli, Perley, \& Harris \markcite{carill194} 1994, which  
shows a complex interaction between
the radio jets and lobes of Cygnus A and the x-ray emitting gas in the
region immediately surrounding the galaxy. The overall x-ray structure has
been interpreted as a cluster-scale cooling-flow, as is usually found in
very rich clusters of galaxies. However,
only four other galaxies in this region have published redshifts close to
that of Cygnus A  
(Spinrad \& Stauffer \markcite{spinrad82} 1982) and thus it has 
appeared that Cygnus A does not
lie in a rich cluster. However, the low galactic latitude (5$^{\circ}$) of Cygnus A 
certainly hampers the detection of a cluster because of the dense, confusing
galactic star field and the relatively high galactic extinction in this
region.

	Given the apparent anomaly of a cooling flow in a poor group
surrounding the most luminous radio galaxy known with a redshift less
than 0.1, we thought this problem was worth another look. Below we report
the detection of an extensive cluster of optical galaxies surrounding Cygnus A 
and briefly discuss its properties as currently known.

\section{Observations and Results}

	A Cousins R-band image of a 22 arcmin field surrounding Cygnus A
was obtained with the KPNO 0.9m telescope. The standard CCD camera on
the 0.9m telescope was used with the 2048X2048 TK2A CCD. The final image was constructed
from three 5 minute exposures. Seeing was 2.2 arcsec for the final
image. The magnitudes were calibrated using about
20 Landolt standards (Landolt \markcite{land92} 1992). Candidate galaxies
on the flattened frame were identified by eye down to a constant peak
surface brightness of 21 mag arcsec$^{-2}$. Very likely this
selection process was incomplete.

	104 of the candidates were then observed using the MX multifiber
spectrograph on the Steward Observatory 2.3-m telescope (Hill \& Lesser
\markcite{hill} 1986) between 12Oct96 and 16Oct96. We used a 400 line mm$^{-1}$ grating covering 3700
to 6950 \AA\ at 8 \AA\ resolution with a UV-flooded Loral 1200$\times$800 pixel
CCD as the detector.

	41 velocities were measured with this system in the velocity
range near the recessional velocity of Cygnus A. With our selection
criteria no redshifts outside this range were found.
We used the IRAF 
cross correlation task,
FXCOR, to derive the velocities, except for Cygnus A itself for which
we used the emission lines. Velocity errors were calculated from the
cross-correlation strength, parameterized by the Tonry \& Davis 
\markcite{tonry} (1979)
R-value. We used the relationship $\Delta V = 280/(1+R)$ km sec$^{-1}$,
derived from redundant MX observations of cluster galaxies (Hill \& 
Oegerle \markcite{ho} 1993).
We have rejected objects with $R < 3$ as unreliable. 

	In table 1, we list the right-ascension and declination (J2000.0), the 
heliocentric velocity and its error estimate, 
and the absolute Cousins R magnitude for each
object from the CCD frame.  Aperture magnitudes were measured within a Gunn-Oke 
aperture (Gunn \& Oke
\markcite{go} 1975), with a metric diameter of 26.2 kpc (assuming $H_0 = 75$
km s$^{-1}$ Mpc$^{-1}$, $q_0=0.1$) and assuming a galactic absorption of 
$A_R =0.81$ (Spinrad \& Stauffer \markcite{spinrad82} 1982) and a 
K-correction of 0.08 magnitudes. The cosmology and velocity given above
for the cluster imply a distance modulus of 37.15. 
 
	After the observations we reexamined the CCD frame using FOCAS
to pick candidate galaxies and then reviewed all the objects from FOCAS
and from our previous visual survey of the image. After rejecting many
close stellar pairs found by FOCAS and some of our previous galaxy
candidates which turn out to be stars, we find 138 non-stellar objects 
brighter than our limiting peak magnitude of which we have measured redshifts 
for 41.  Thus we have redshifts for about 30\% of these objects. Presumably,
the remaining objects are a combination of cluster members, background
objects and probably some stars.

\section{Analysis and Discussion}

	In figure 1, we show the velocity histogram for the 41 galaxies 
in table 1. The central biweight velocity is 18,873${^{+322}_{-332}}$ 
 km s$^{-1}$ and the biweight
scale (dispersion) is 1581${^{+286}_{-197}}~\rm km\ s^{-1}$ 
corrected to the restframe of the cluster(see Beers \etal\ 
\markcite{beers90}1990 for a description of the biweight estimators). 
This is a large, but not unprecedented
dispersion (Oegerle, Hill \& Fitchett \markcite{OHF}1995) for a very
rich cluster. Most such high-dispersion clusters are made up of two or more subclusters
and figure 1 hints at a bimodal distribution. 

We applied a battery of 1-D, 2-D, and 3-D statistical tests (see Pinkney \etal\ 
\markcite{pink96} 1996) 
in order to determine the form and amount of both velocity and spatial substructure.
6/8 of the kurtosis tests performed returned a significant statistic ($\geq 90\%$ probability), 
while none of the tests sensitive to skewness were significant. Thus the velocity
distribution in figure 1 has a {\it heavier tail} than expected for a Gaussian 
distribution.  This 1-D substructure is a result of the secondary peak at 
the velocity of Cygnus A.  


In figure 2, we show a velocity-coded diagram indicating the spatial
location of the member galaxies.  Certainly there appears to be no 
simple segregation of velocities.  
The Dressler-Shectman test [DS] (Dressler \& Shectman \markcite{ds88} 1988) 
gives a cumulative delta of only 45.5, 
significantly different from a random distribution of galaxy velocities 
and positions at the 82\% level (no evidence for a spatially segregated 
subcluster).  The DS test, however, is insensitive to substructure if
the cores of the two subclusters are too close together in angular or
redshift space as could be the case here. 

     Of the three 3-d statistical tests we applied to the 
sample, only the $\alpha$-test (West \& Bothun \markcite{wb90} 1990) 
returned a significant result (at the 98\% level).  This test 
is a measure of how much the centroid of the galaxy distribution shifts
due to correlations between the local kinematics (of the $\sqrt N$ nearest
neighbors) and the projected galaxy distribution.  This test is also 
insensitive to a merger with a small projection angle. We suspect
that the significant result arises from the grouping of 
lower-velocity galaxies to 
the NW of the centroid in figure 2, which would have a small
local velocity dispersion and large offset from the dynamical centroid. 
Note that this galaxy grouping does not include Cygnus A itself. 

	In figure 3, we show the adaptively smoothed distribution of
galaxies from table 1 overlayed on the ROSAT PSPC image of the Cygnus cluster.
One can see that the galaxy distribution follows the diffuse x-rays
fairly well. However, the bright clump containing Cygnus A lies offset
from both the x-rays and the galaxies. The velocity of Cygnus A (16811
km s$^{-1}$ 
is offset by 2039 km s$^{-1}$  from the cluster peak (figure 1). 
Thus there appears to be a very rich
cluster in the direction of Cygnus A but the radio galaxy does not
appear to be at its center.

	Using a Schechter function with $M_{R}^{*}=-22$, $\alpha=-1.25$
 and a King law with $r_c=250$ kpc (to estimate the
number of galaxies missed due to the field of view of the CCD), we can set 
a lower limit to the Abell richness class of 1. If we are incomplete
by about a factor of two for cluster members as we suspect, the cluster 
would be richness class 3. If all 138 objects we have found still 
without redshifts were cluster members the richness could be as high as 4
but this seems unlikely since some objects are likely to be foreground or
background galaxies and, based on past experience, some are probably stars.

	Such a rich cluster in the vicinity of Cygnus A both in
celestial and velocity space seems unlikely to occur by chance. Other
such examples of high velocity dispersion clusters have been explained
as cluster-cluster mergers in progress. The dense core of hot
gas surrounding Cygnus A, often called a ``cooling flow'', is typical of
the cores of relaxed, very rich clusters. The offset position of Cygnus A
suggests that the dense core may be a remnant of a rich cluster
which is taking part in a merger with another system. 
Alternatively, Cygnus A could be the dominant galaxy in a poor cluster
merging with a richer system.
Such a merger or
collision, if it is taking part supersonically, could compress and heat
the gas in the surviving core. This would in turn enhance and confine the
radio emitting plasma and could be partly responsible for the very high
radio luminosity of Cygnus A. Such a process might also be capable of
stimulating the AGN itself.   

	Further measured velocities are necessary for the region
around Cygnus A to clarify whether this system is actually a cluster-cluster
merger in progress and if so what sort of a merger scenario is consistent
with the system. Many fainter galaxies, for which we have not been able to 
measure redshifts, lie in the vicinity of the radio galaxy. Also we have only
probed about about 700 kpc from the center of Cygnus A in projection. Thus
many more cluster members are likely to lie outside our search radius.

\section{Conclusions}

	The Cygnus A radio galaxy lies within a rich, possibly very rich, 
high velocity
dispersion cluster. It appears to be offset from the cluster center
both in physical space and in velocity. The existence of a very powerful 
radio galaxy/AGN in a dense, hot gaseous core is at least very interesting 
and could well be a key to the origin of such systems.

	We  thank Bill Oegerle for comments on the text. M.J.L.
acknowledges partial support from NSF Grant AST-9317596.

\clearpage

\begin{deluxetable}{cccccc}
\small
\tablewidth{250pt}
\tablenum{1}
\tablecaption{Cygnus-A cluster velocities} 
\tablehead{
\colhead{RA}        &
\colhead{DEC}     & \colhead{$\rm V_H$}       &
\colhead{$\Delta V$}  & \colhead{$M_R$}  }
\startdata
19:58:33.1 & 40:42:47 & 18161 & 57 & -20.6\nl
19:58:33.7 & 40:52:20 & 19295 & 41 & -21.2\nl
19:58:40.3 & 40:50:01 & 15966 & 55 & -21.2\nl
19:58:41.1 & 40:54:44 & 19359 & 61 & -21.0\nl
19:58:42.8 & 40:45:28 & 17095 & 49 & -21.2\nl
19:58:45.5 & 40:45:38 & 19031 & 48 & -22.7\nl
19:58:45.8 & 40:48:51 & 19924 & 40 & -21.5\nl
19:58:48.1 & 40:36:55 & 18599 & 57 & -20.4\nl
19:58:56.8 & 40:48:51 & 16119 & 44 & -21.2\nl
19:58:59.2 & 40:53:19 & 17070 & 61 & -20.9\nl
19:59:01.0 & 40:51:07 & 19479 & 42 & -22.4\nl
19:59:06.3 & 40:36:20 & 20242 & 30 & -21.6\nl
19:59:06.7 & 40:44:03 & 20219 & 31 & -21.9\nl
19:59:08.1 & 40:54:13 & 16495 & 50 & -22.6\nl
19:59:11.3 & 40:51:25 & 18256 & 34 & -23.0\nl
19:59:13.4 & 40:50:05 & 19084 & 70 & -21.7\nl
19:59:14.1 & 40:48:15 & 15936 & 44 & -21.2\nl
19:59:16.3 & 40:50:07 & 15563 & 60 & -21.4\nl
19:59:26.1 & 40:44:08 & 14805 & 23 & -21.5\nl
19:59:28.3 & 40:44:02 & 16811 & 20 & -23.4\nl
19:59:32.6 & 40:49:25 & 19266 & 51 & -20.8\nl
19:59:33.8 & 40:52:17 & 19738 & 64 & -21.3\nl
19:59:33.7 & 40:47:09 & 19630 & 33 & -21.6\tablebreak
19:59:34.0 & 40:47:35 & 20071 & 38 & -21.4\nl
19:59:35.8 & 40:50:23 & 19972 & 22 & -22.4\nl
19:59:36.4 & 40:42:06 & 18239 & 70 & -21.2\nl
19:59:37.3 & 40:36:47 & 18924 & 36 & -21.2\nl
19:59:40.3 & 40:47:15 & 21283 & 44 & -21.7\nl
19:59:41.5 & 40:42:04 & 19193 & 68 & -20.7\nl
19:59:44.2 & 40:47:01 & 17039 & 32 & -22.0\nl
19:59:49.4 & 40:44:37 & 20706 & 35 & -21.2\nl
19:59:55.0 & 40:40:56 & 19652 & 70 & -20.9\nl
19:59:55.3 & 40:45:20 & 19654 & 27 & -23.1\nl
19:59:55.6 & 40:43:56 & 20090 & 70 & -20.9\nl
19:59:56.1 & 40:38:27 & 19780 & 34 & -21.9\nl
20:00:02.1 & 40:45:58 & 18661 & 57 & -21.0\nl
20:00:03.4 & 40:40:42 & 18819 & 39 & -22.4\nl
20:00:06.0 & 40:36:36 & 18387 & 38 & -22.2\nl
20:00:07.7 & 40:46:03 & 17726 & 50 & -20.4\nl
20:00:14.7 & 40:50:13 & 19454 & 37 & -21.3\nl
20:00:17.4 & 40:51:17 & 23355 & 64 & -21.2\nl
\enddata
\end{deluxetable}

\begin{figure}[p]
\plotone{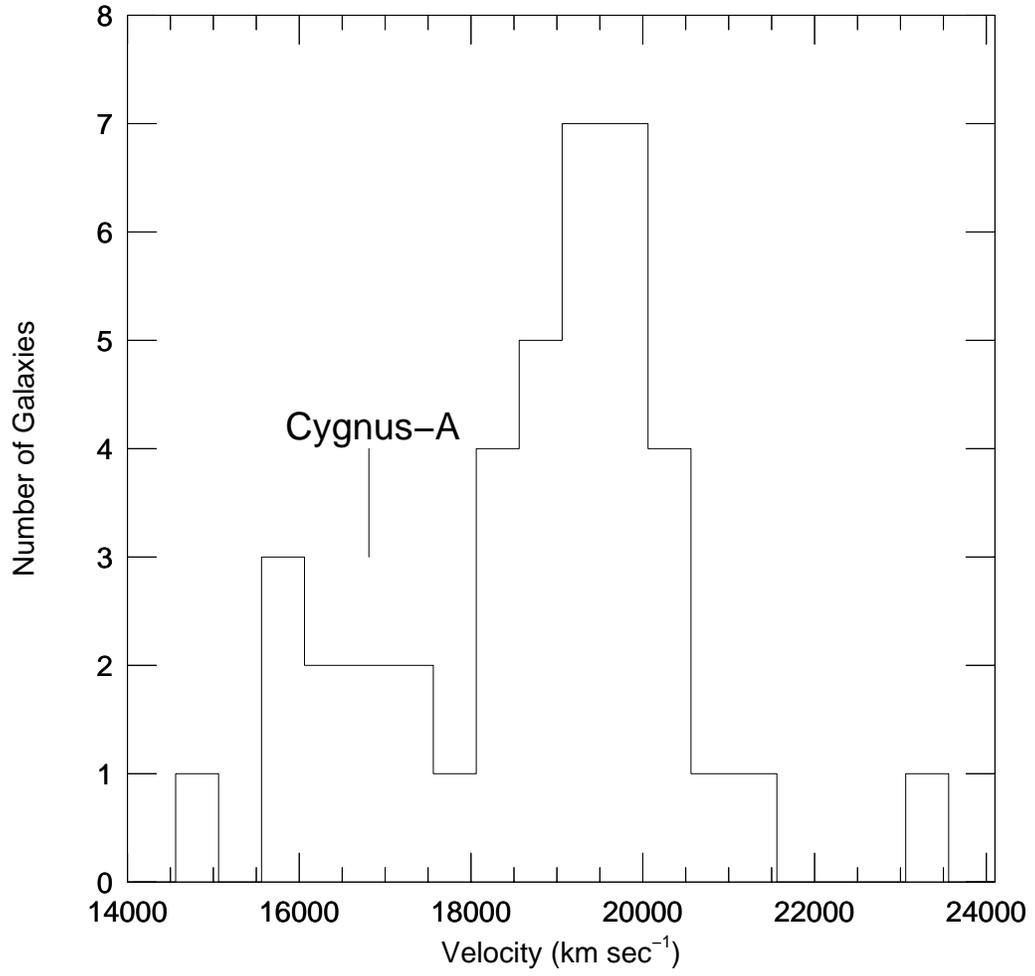}
\caption
{Radial velocity histogram from Table 1}
\end{figure}

\clearpage
\begin{figure}[p]
\plotone{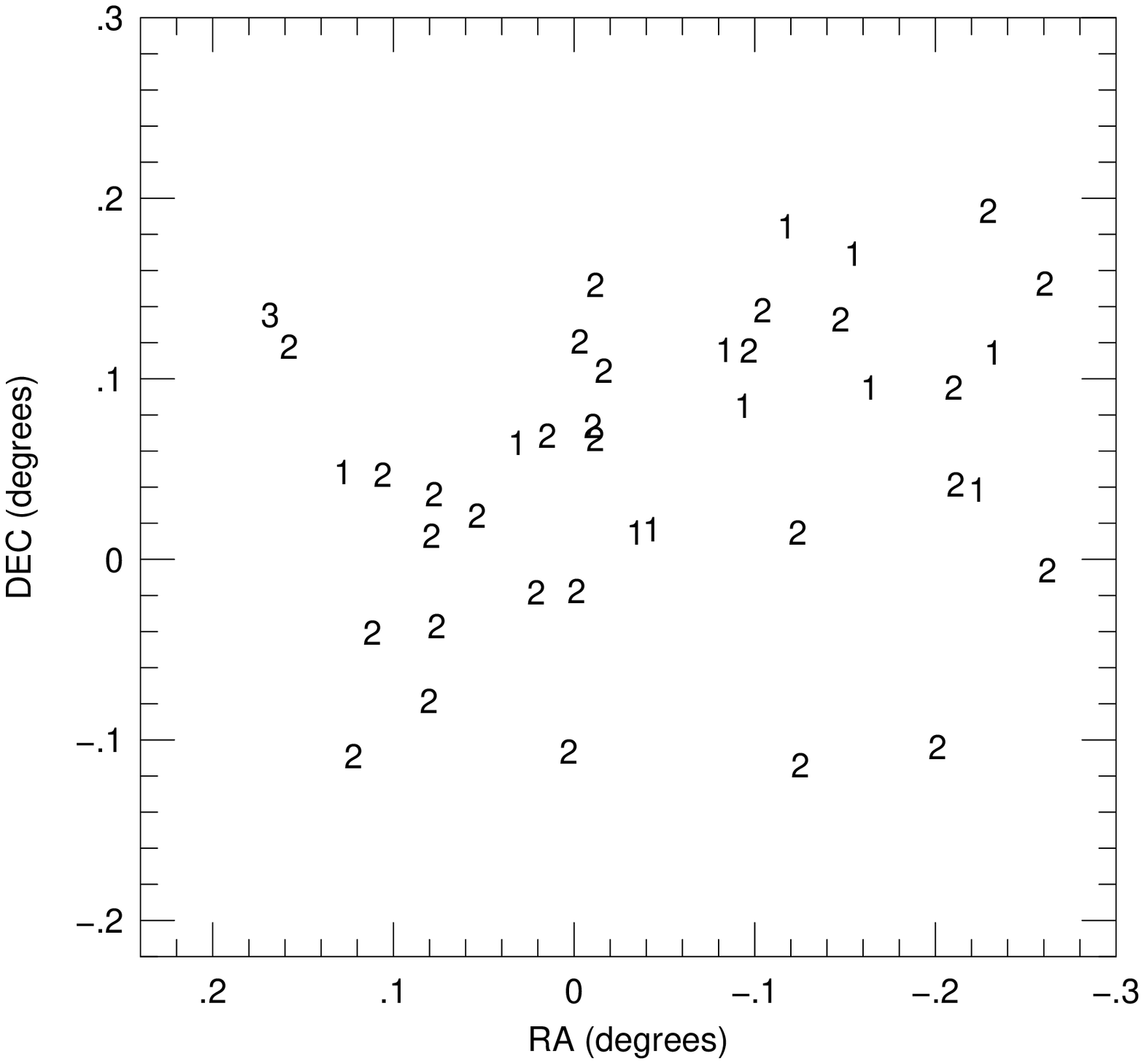}
\caption
{Spatial distribution of galaxy velocities from Table 1,
binned in the velocity ranges, 1=$<18000$, 2=$18000-22000$, 3=$>22000$.
Cygnus A is the point labelled `1' nearest coordinate (0,0).}
\end{figure}

\clearpage
\begin{figure}[p]
\plotone{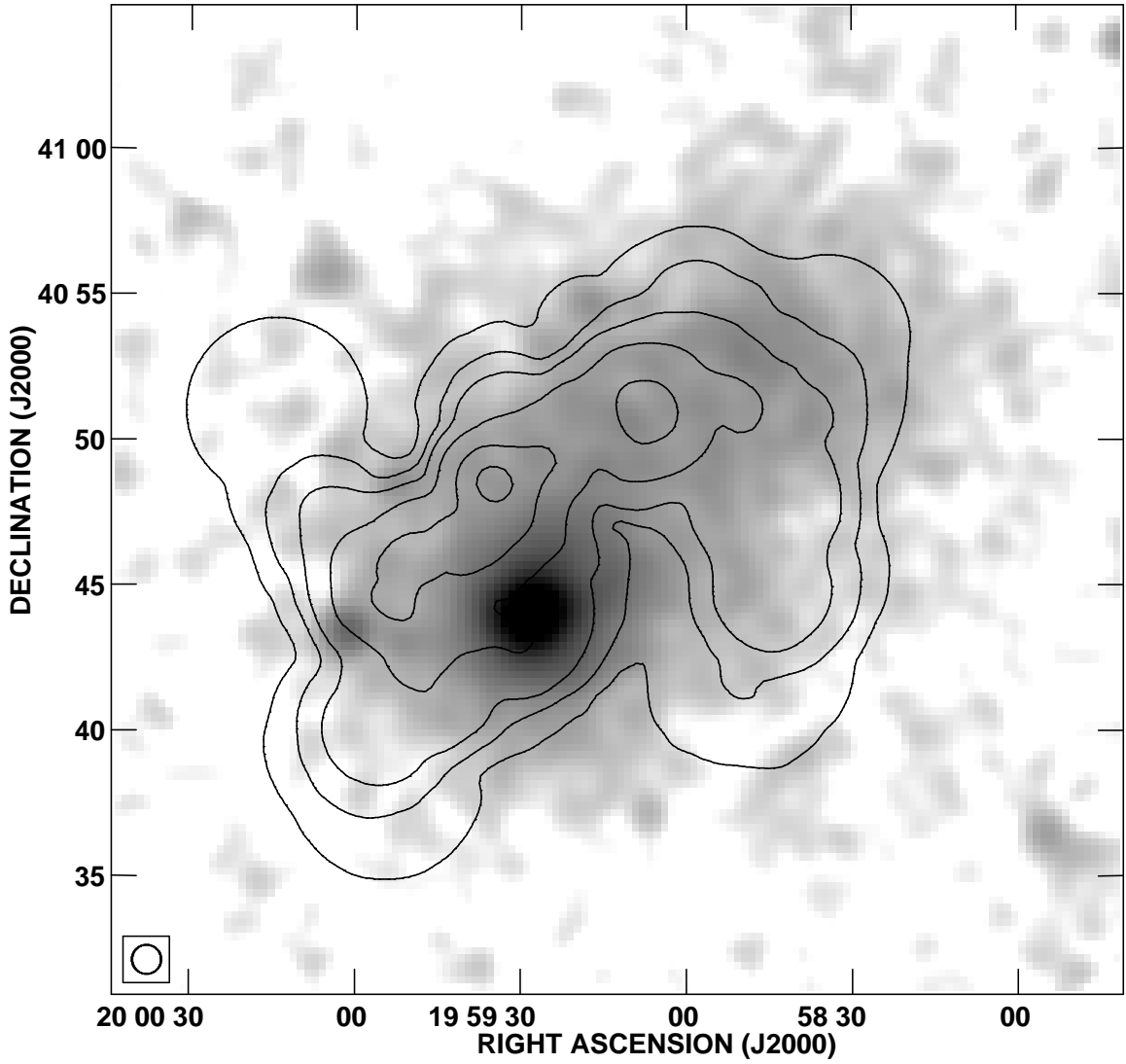}
\caption
{The adaptively smoothed galaxy distribution (contours)
superimposed
on the ROSAT PSPC x-ray image (grey-scale). The galaxy density contours
are 1,2,3,5,8,$10 \times 0.03$ galaxies per square arcmin. The image is
centered on the dynamical centroid. Cygnus A is located at the x-ray
peak of the grey-scale image.}
\end{figure}

\end{document}